\documentclass[amssymb,prl,aps,twocolumn,amsmath,showpacs]{revtex4}

\usepackage{graphicx}

\begin{document}

\title{Magnetic-field-dependent quasiparticle energy relaxation in
mesoscopic wires}
\author{A.\ Anthore, F.\ Pierre, H. Pothier, and D. Esteve}
\affiliation{Service de Physique de l'Etat Condens\'{e}, Direction des Sciences\\
de la Mati\`{e}re, CEA-Saclay, 91191 Gif-sur-Yvette, France}
\date{January 6, 2003}

\begin{abstract}
In order to find out if magnetic impurities can mediate interactions between
quasiparticles in metals, we have measured the effect of a magnetic field $B$
on the energy distribution function $f(E)$ of quasiparticles in two silver
wires driven out-of-equilibrium by a bias voltage $U$. In a sample showing
sharp distributions at $B=0$, no magnetic field effect is found, whereas in
the other sample, rounded distributions at low magnetic field get sharper as
$B$ is increased, with a characteristic field proportional to $U$.
Comparison is made with recent calculations of the effect of
magnetic-impurities-mediated interactions taking into account Kondo physics.
\end{abstract}


\maketitle

The understanding of the phenomena which, at low temperature, limit the
extent of quantum coherence in electronic transport and allow the
quasiparticles to exchange energy is presently an important issue in
mesoscopic physics. There is indeed a discrepancy between the theory \cite%
{AA}, which predicts that Coulomb interactions provide the dominant
mechanism for decoherence and for energy exchange, and measurements of the
coherence time \cite{MW,Lin} or of energy exchange rates \cite%
{relax,relaxAg,Pesc,Fred} in numerous metallic samples. This discrepancy has
been attributed either to a flaw in the theory \cite{MW}, or to the presence
in these samples of other mechanisms involving the scattering of electrons
by undetected two-level systems or magnetic impurities. It has been indeed
recently predicted that even a minute concentration of such scatterers would
result in sizeable energy exchange if the Kondo effect occurs \cite%
{Glazman,Georg,Kroha}. Whereas the limitation of quantum coherence by the
Kondo effect is widely known \cite{haes}, its efficiency for mediating
energy exchange between quasiparticles had not been anticipated. In the case
of magnetic impurities, a significant weakening of this effective
electron-electron interaction is furthermore predicted when a large magnetic
field is applied \cite{Georg2}. In order to test these new predictions and
more generally to understand inelastic processes in mesoscopic conductors,
we have investigated the magnetic field dependence of the energy exchange
rate in mesoscopic wires.

The samples are wires connected to reservoirs biased at potentials $0$ and $U
$ (see Fig.$~$1)$.$ The energy distribution function in the middle of the
wire, $f(E)$, depends on the ratio of the typical interaction time $\tau
_{int}$ and the diffusion time of quasiparticles $\tau _{D}=L^{2}/D.$ If $%
\tau _{int}\gg \tau _{D},$ interactions can be neglected and $f(E)$ is the
average of the Fermi functions in both reservoirs, which have
electrochemical potentials shifted by $eU.$ In the experimental situation
where $k_{B}T\ll eU,$ $f(E)$ is then a two-step function. In the opposite
limit $\tau _{int}\ll \tau _{D},$ local equilibrium is achieved at each
coordinate along the wire, and $f(E)$ is a Fermi function at a temperature
given by the balance between Joule heating and electronic heat conductivity
to the reservoirs: this is the ``hot-electron'' regime \cite{Andy}. The
intermediate regime is of interest for experiments because the precise shape
of $f(E)$ and its dependence on $U$ are characteristic of the interaction
rate and of its energy dependence \cite{relax}.

\begin{figure}[tbp]
\includegraphics[width=3.1in]{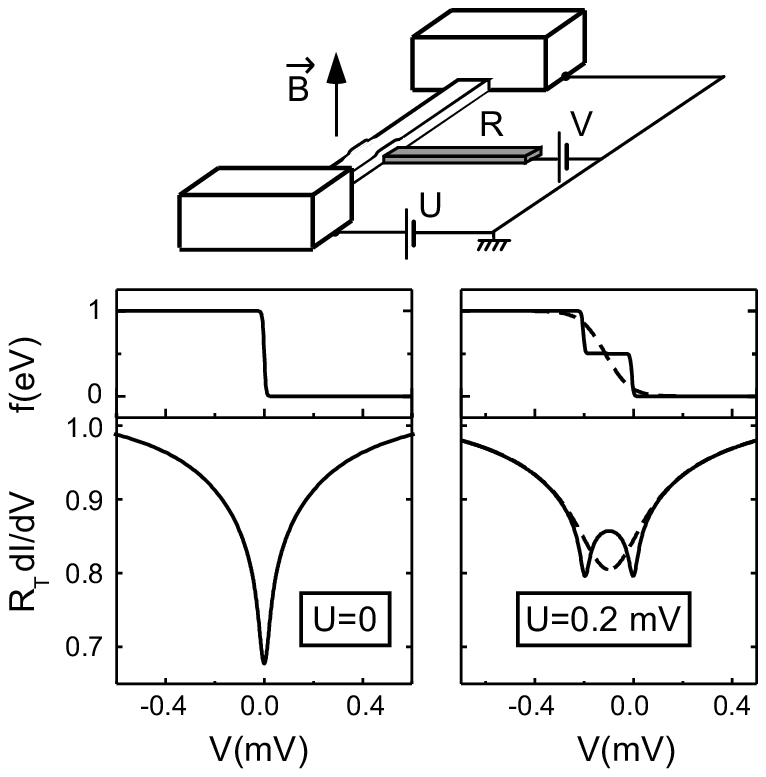}
\caption{Top: layout of the experiment: a wire is connected to two large electrodes biased at potentials $0$ and
$U$. A resistive probe electrode (in grey) forms a tunnel junction with the wire. At equilibrium $(U=0)$, the
differential conductance $dI/dV(V)$ of this junction displays a dip at zero bias, due to Coulomb blockade of
tunneling (left). When the quasiparticles of the wire are driven out-of-equilibrium by a finite voltage $U$
(right)$,$ their energy distribution function $f(E)$ depends on the interaction rate between quasiparticles. In
the absence of interactions, $f(E)$ is a two-step function and $dI/dV(V)$ presents two dips (solid lines). With
strong interactions, $f(E)$ is rounded, and $dI/dV(V)$ presents a broad dip (dashed lines).} \label{fig1}
\end{figure}

At zero magnetic field, the distribution function $f(E)$ can be inferred
from the differential conductance $dI/dV(V)$ of a tunnel junction between
the central part of the wire and a superconducting (aluminum) probe
electrode biased at potential $V$ \cite{relax}$.$ In magnetic fields larger
than the critical field $B_{c}\sim 0.1~{\rm T}$ of the superconducting
electrode, another method is required. Here, we have taken advantage of the
nonlinearity of the current-voltage characteristic of a tunnel junction
placed in series with a resistance $R.$ When both electrodes of the junction
are in the normal state and at thermal equilibrium, the differential
conductance $dI/dV(V)$ displays a dip at $V=0$ (see Fig.~1), due to the
Coulomb blockade of tunneling \cite{environment}. Assuming that the two
electrodes have different distribution functions $f$ and $f_{{\rm ref}},$
the differential conductance reads:%
\begin{eqnarray} \frac{dI}{dV}(V)&=&\frac{1}{R_{{\rm T}}}\int dE~f(E)~\int d\varepsilon
~P(\varepsilon ) \label{dIdV3}\\
&\times& \frac{\partial }{\partial E}((f_{{\rm ref}%
}(E+eV+\varepsilon )-f_{{\rm ref}}(E+eV-\varepsilon ))\nonumber
\end{eqnarray}
where $R_{{\rm T}}$ is the tunnel resistance of the junction, and $%
P(\varepsilon )=\int \frac{dt}{2\pi \hbar }~{\rm e}^{J(t)+i\varepsilon
t/\hbar }~$the probability for an electron to tunnel through the barrier
while releasing to the environment an energy $\varepsilon $, $J(t)=\int
\frac{d\omega }{\omega }~\frac{2%
\mathop{\rm Re}%
[Z(\omega )]}{R_{K}~}~\frac{{\rm e}^{-i\omega t}-1}{1-{\rm e}^{-\hbar \omega
/k_{B}T}}$ with $Z(\omega )=1/(1/R+jC\omega ),$ $C$ the junction
capacitance, $R_{K}=h/e^{2}\approx 25.8~{\rm k\Omega }$ the resistance
quantum and $T$ the environment temperature. In the case where the
distribution function $f(E)$ presents two steps, as in Fig. 1, and $f_{{\rm %
ref}}$ is a Fermi function at temperature $T$, one obtains, by linearity,
two dips in $dI/dV(V)$ at $V=0$ and $V=-U.$ In contrast, in the hot electron
regime, $dI/dV(V)$ displays a broad dip centered at $V=-U/2$ (see Fig. 1).
In the experiments, a large series impedance at the relevant frequencies (up
to about 50$~$GHz) was obtained by designing the probe electrode as a long,
narrow and thin aluminum electrode (25$\mu $m$\times $150nm$\times $12nm),
which presents a resistance $R\sim 1.5~{\rm k\Omega }$ in the normal state.

We present here the results obtained on two silver samples in which the
distribution functions found at $B=0$ were extremely different. The samples
were obtained from nominally five-nines-purity (99.999\%, sample \#1) and
six-nines-purity (99.9999\%, sample \#2) source material. For both wires,
the length and cross-section area are $L=20~{\rm \mu m,}$ $S=100~{\rm nm}$ $%
\times 48~{\rm nm}.$ The diffusion constants $D=196$ and $215~{\rm cm}^{2}%
{\rm /s}$ respectively, were deduced from the low temperature resistance.
The tunnel resistances $R_{{\rm T}}$ ($167~$and $102~{\rm k\Omega )}$ and
the capacitances $C$ (0.8 and 0.9~fF) of the junctions, as well as the
environment resistances $R$ (1.34 and 1.65 ${\rm k\Omega }$), were obtained
from fits with Eq.~(\ref{dIdV3}) of $dI/dV(V)$ measured at $B=0.3~${\rm T}
and $U=0$. We have checked that these curves do not change with $B$ when $%
B>B_{c}$.

\begin{figure}[tbp]
\includegraphics[width=3.4in]{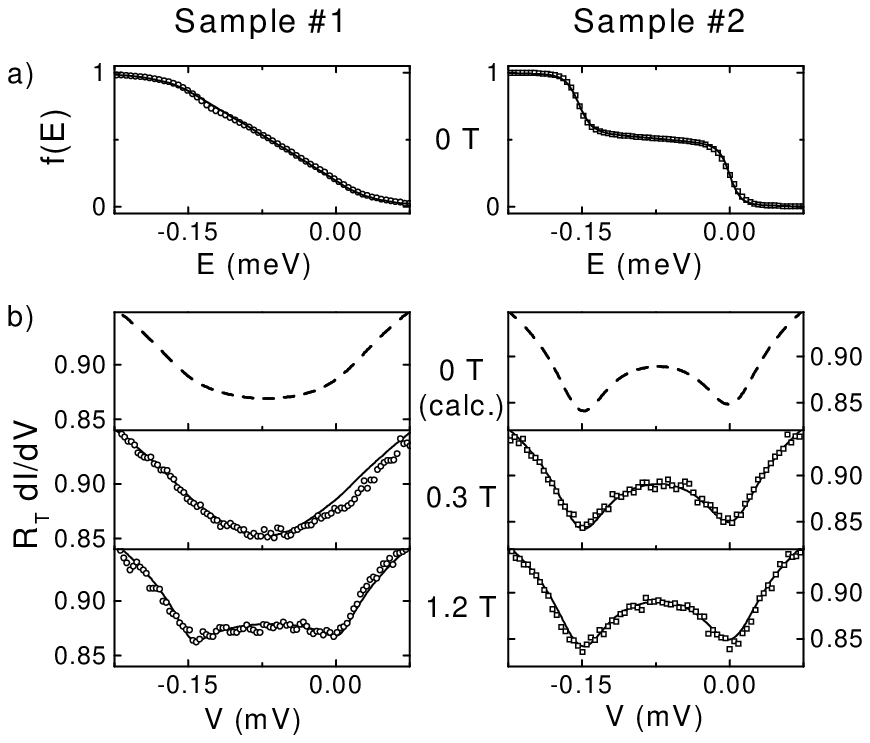}
\caption{a) Symbols: distribution functions $f(E)$ at $U=0.15~$mV and zero
magnetic field in samples \#1\ and \#2, obtained by deconvolution of $%
dI/dV(V)$ with the probe electrode in the superconducting state. Solid lines: fits with theory including the
effect of Kondo impurities (see text). b) Dashed line: calculated Coulomb blockade signal $dI/dV(V)$ using the
measured $f(E)$ at $B=0.$ Symbols: measured $dI/dV(V)$ at $U=0.15~$mV, with $%
B=0.3$ and $1.2~$T, the probe electrode being in the resistive state. Solid line: fits with theory.}
\label{fig2}
\end{figure}

At low magnetic field and low temperature, the probe electrode is
superconducting. Its impedance is purely imaginary at frequencies lower than
$2\Delta /h$ \cite{tinkham}. It results that for $eV\in \left[ -3\Delta
+U,3\Delta \right] $ Coulomb blockade only leads to a reduction of the
differential conductance, which is multiplied by a factor $\exp
(-\int_{0}^{\infty }\frac{d\omega }{\omega }~\frac{2%
\mathop{\rm Re}%
[Z(\omega )]}{R_{K}~})$ $\sim 0.9$. Numerical deconvolution of $dI/dV(V)$ is
therefore possible, and the distribution functions obtained at $U=0.15$ mV
are presented in the top of Fig. 2 for both samples. Whereas $f(E)$ is close
to a double-step function in sample \#2, it is much more rounded in sample
\#1, indicating that the energy exchange rate is much larger in the latter,
since the diffusion times are very similar ($\tau _{D}=L^{2}/D\simeq 20~{\rm %
ns)}$. In the bottom of Fig.~2, we plot the calculated $R_{T}dI/dV(V)$ using
formula (\ref{dIdV3}) with $f(E)$ the distribution function measured at $B=0$
(dashed curves), and present the measured curves for $B=0.3~{\rm T}$ and $%
B=1.2~{\rm T}$ (symbols){\rm \ }\cite{debloc}$.$ In sample \#2, the magnetic
field has no visible effect. Note however that the distribution functions
are so close to a double-step that the experiment is not sensitive enough to
detect a possible slight reduction of the energy exchange rate with $B$. In
contrast, in sample \#1, the rounded dip at zero field is replaced at $1.2~%
{\rm T}$ by a double-dip, showing that the energy exchange rate has been
reduced. Figure 3 shows the evolution of $dI/dV(V)$ with magnetic field,
from $0.3~{\rm T}$ to $1.5~{\rm T}$ by steps of $0.3~{\rm T,}$ for $U=0.1,$
0.2 and $0.3{\rm ~mV}$. A similar behavior is observed at all values of $U$:
the low-field broad conductance dip at $B=0.3~{\rm T}$ tends to be replaced
at large fields by a double-dip structure. In particular, the crossover
field at which $dI/dV(V)$ is nearly constant over a broad voltage range is $%
0.6~{\rm T}$ at $U=0.1{\rm ~mV,}$ $0.9~{\rm T}$ at $U=0.15{\rm ~mV}$ (not
shown), $1.2~{\rm T}$ at $U=0.2{\rm ~mV,}$ and $1.5~{\rm T}$ at $U=0.25{\rm %
~mV}$ (not shown)${\rm ,}$ hence presenting a linear increase with $U.$ The
comparison of the raw data on sample \#1 and sample \#2 in Fig.~2 already
allows to conclude that sample \#1 presents an extra interaction which can
be strongly reduced by applying a magnetic field.

\begin{figure}[tbp]
\includegraphics[width=3.4in]{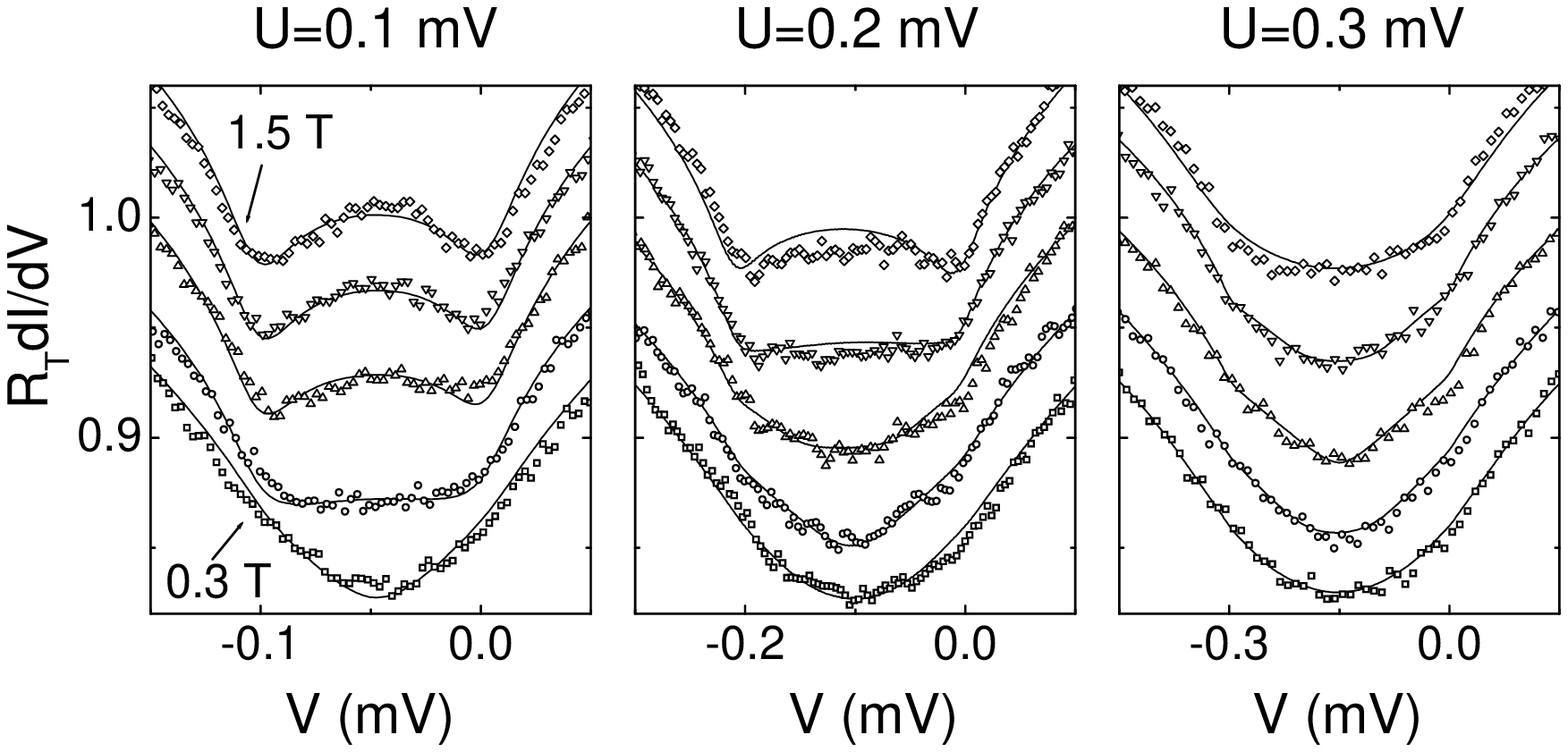}
\caption{Magnetic field effect in sample \#1: differential conductance $%
dI/dV(V)$ at $U=0.1,$ $0.2$ and $0.3~$mV, for $B$ ranging (from bottom to top) from 0.3 to 1.5~T by steps of
0.3~T. Successive curves have been vertically offset by steps of 0.033, for clarity.} \label{fig3}
\end{figure}

We now compare the experimental data with theoretical predictions. The distribution function is calculated by
solving the stationary Boltzmann equation in the diffusive regime~\cite{Kozub,Nagaev}:
\begin{equation}
\frac{1}{\tau _{D}}\frac{\partial ^{2}f\left( x,E\right) }{\partial x^{2}}=-%
{\cal I}_{{\rm coll}}^{{\rm in}}\left( x,E,\left\{ f\right\} \right) +{\cal I%
}_{{\rm coll}}^{{\rm out}}\left( x,E,\left\{ f\right\} \right)
\label{Boltzmann}
\end{equation}%
where ${\cal I}_{{\rm coll}}^{{\rm in}}$ and ${\cal I}_{{\rm coll}}^{{\rm out%
}}$ are the rates at which quasiparticles are scattered in and out of a
state at energy $E$ by inelastic processes. Assuming that the dominant
inelastic process is a two-quasiparticle interaction which is local on the
scale of variations of the distribution function,
\begin{equation}
{\cal I}_{{\rm coll}}^{{\rm in}}=\int {\rm d}\varepsilon {\rm d}E^{\prime
}K\left( \varepsilon \right) f_{E+\varepsilon }^{x}(1-f_{E}^{x})f_{E^{\prime
}}^{x}(1-f_{E^{^{\prime }}-\varepsilon }^{x})  \label{Iout}
\end{equation}%
where the shorthand $f_{E}^{x}$ stands for $f\left( x,E\right) .$ The
out-collision term ${\cal I}_{{\rm coll}}^{{\rm out}}$ has a similar form.
The kernel function $K\left( \varepsilon \right) $ is proportional to the
averaged squared interaction between two quasiparticles exchanging an energy
$\varepsilon .$ Coulomb interactions lead, in diffusive wires, to $%
K(\varepsilon )=\kappa /\varepsilon ^{3/2}$ \cite{AA}, where $\kappa =(\pi
\sqrt{\frac{D}{2}}\hbar ^{3/2}\nu _{F}S)^{-1}$ with $\nu _{F}$ the density
of states at the Fermi level \cite{Kamenev}. The $B=0$ data for sample \#2
can be well fit using this term with $\kappa =0.12$\noindent ~${\rm ns}^{-1}%
{\rm meV}^{-1/2}$, of the same order of magnitude as the theoretical value $%
0.07$ ${\rm ns}^{-1}{\rm meV}^{-1/2}$ \cite{prefactor}, and a term of lesser
importance describing phonon emission \cite{eph}. The $B=0$ data for sample
\#1 can be fit similarly, with $\kappa =2.4$\noindent\ ${\rm ns}^{-1}{\rm meV%
}^{-1/2}$, however the reduction of the energy exchange rate with $B$
indicates that an extra process is present at $B=0$. We have in the
following fixed $\kappa $ to the best fit value obtained for the large
field, low $U$ data, where the $B-$dependent interaction has essentially
vanished: $\kappa =0.5$\noindent $~{\rm ns}^{-1}{\rm meV}^{-1/2}$ \cite%
{prefactor}. The remaining part of the energy exchange rate was fit with the
theory of {G\"{o}ppert }{\em et al. }\cite{Georg2,fitgeorg}, which accounts
for the effective interaction in the presence of a concentration $c$ of spin-%
$\frac{1}{2}$ impurities, with Kondo temperature $T_{K}$, gyromagnetic
factor $g$, and coupling constant $J$ between quasiparticles and magnetic
impurities. The Kondo effect is included in this calculation, under the
assumption that the distribution functions are not too sharp, leading to a
renormalization of $J$ depending on the distribution function itself. The
corresponding inelastic integral can also be written in the form of Eq.~(\ref%
{Iout}), but with a $K\left( \varepsilon \right) $ function depending on the
energies $E$ and $E^{\prime }$ and on $f$. At zero magnetic field, the
effect of this term on $f(E)$ is similar to that of a phenomenological
kernel $K\left( \varepsilon \right) \propto 1/\varepsilon ^{2}$ as found in %
\cite{relax}. For compatibility with phase-coherence time measurements (see below), the Kondo temperature was
fixed at $T_{K}=40~{\rm mK,}$ which is the Kondo temperature of Mn in Ag. As shown by solid lines in Fig. 2 and
in Fig. 3, the data can be accurately reproduced using $c=17~{\rm ppm},$ $g=2.9$ and ${\nu J=}0.08$ \cite{asym}.
Note however that according to material analysis of the silver batch used to fabricate sample \#1, no magnetic
impurity was present in the source at the level of 1~ppm. Since in some samples made out of the same batch the
intensity of the energy exchange rate measured at $B=0$ was found to be up to 4 times smaller, pollution of the
sample during fabrication might, at least partly, explain this large impurity concentration.

The impurity concentration deduced from the fits of $f(E)$ must be further
compared with the one obtained from the analysis of measurements of the
phase coherence time in long wires fabricated previously with the same
source materials. We have extracted the phase coherence time $\tau _{\varphi
}$ from the magnetoresistance of wires much longer than the phase coherence
length, using the weak localization theory. In samples made of 6N purity Ag,
$\tau _{\varphi }(T)=A/T^{2/3}$ from $1~{\rm K}$ down to $40~{\rm mK}$, with
$A=2.25~{\rm ns\ K}^{2/3},$ in reasonable agreement with the theory of
Coulomb interactions in disordered wires ($A_{{\rm theory}}=3.00~{\rm ns\ K}%
^{2/3}$). At $T=40~{\rm mK}$, $\tau _{\varphi }=18~{\rm ns}$. In samples
made of 5N silver, $\tau _{\varphi }(T)$ does not vary between $T=200~{\rm mK%
}$ and $40~{\rm mK}$, where we find $\tau _{\varphi }=2~{\rm ns}$. This
behavior can be attributed to the presence of magnetic impurities, with
concentration $c$, spin $s$ and Kondo temperature $T_{K}$, which lead to a
spin-flip rate described by \cite{Nagaoka,haes} $\gamma _{sf}(T)$\noindent $%
~=(c/\pi \hbar \nu )~\pi ^{2}s(s+1)/(\pi ^{2}s(s+1)+\ln ^{2}(T/T_{K}))$. The
resulting phase coherence time $\tau _{\varphi }(T)=1/(T^{2/3}/A+2\gamma
_{sf}(T))$ shows very little variation between $40~{\rm mK}$ and $200~{\rm mK%
}$ and describes precisely the experimental data for $c=0.1~{\rm ppm}$, $%
T_{K}=40~{\rm mK}$, $s=1/2$ and $A=1.95~{\rm ns\ K}^{2/3}$ ($A_{theory}=2.6~%
{\rm ns\ K}^{2/3}$). This value of $c,$ compatible with the nominal source
purity, is smaller by two orders of magnitude than the value obtained from
the fits of energy exchange data on sample \#1. A similar set of results was
also obtained with Cu samples, a material in which the oxyde at the surface
of the films was found to cause dephasing at low temperature \cite{Vranken}.
Data on energy exchange \cite{moriond} could also be fit with the theory of G%
\"{o}ppert {\em et al }\cite{Georg2}{\em , }using $T_{K}=300~{\rm mK},$ $%
c=4.8~{\rm ppm,}$ $g=2.3$, $\nu J=0.1$, on top of a Coulombic term with
intensity $\kappa =0.5~{\rm ns}^{-1}{\rm meV}^{-1/2}$ \cite{fitgeorg}. This
result gives evidence that the anomalous interactions observed in many Cu
wires at $B=0$ \cite{relax,Fred} are also due to magnetic impurities. Here
also, measurements of the phase coherence time \cite{Fred} are explained by
significantly smaller impurity concentrations ($\sim $0.3 ppm). This
repeated discrepancy on the concentrations deduced from the two types of
measurements remains an open problem. From an experimental point of view, a
more quantitative test of theory could be obtained in samples with added,
identified magnetic impurities at a known concentration \cite{pbjunctions}.

To conclude, we have found that anomalous energy exchange rates between
quasiparticles were strongly reduced by the application of a magnetic field.
Moreover, the energy and magnetic field dependence of the exchange rate can
be accurately accounted for by the presence of a small concentration of
Kondo magnetic impurities \cite{Georg2}. It is worthwhile to compare this
result with recent measurements on Aharonov-Bohm rings, which show that the
small phase-coherence times found at $B=0$ were increased in a finite
magnetic field \cite{AB}. All these measurements indicate that the presence
of very dilute magnetic impurities is a very plausible candidate to explain
both extra dephasing and extra energy exchange observed in many mesoscopic
samples.

We acknowledge the technical help of P. Orfila, fruitful discussions and
correspondence with G. G\"{o}ppert, H. Grabert and N. Birge, and permanent
input from M. Devoret, P. Joyez, C. Urbina and D. Vion.


\begin{thebibliography}{}
\bibitem{debloc} Because the signal to noise ratio is not large enough,
numerical deconvolution of the $dI/dV(V)$ curves proved to be unreliable. In
particular, the preliminary $f(E)$ curves in \cite{moriond} and in \cite%
{Georg2} were artificially rounded. We prefer to present here raw data.
\bibitem{prefactor} In other experiments with silver wires \cite{relaxAg},
the fit value of $\kappa $ was found to be between 3 and 10 times the
theoretical one \cite{Fred}. This scatter in the intensity of the
interaction is presently not understood.
\bibitem{fitgeorg} In Ref. \cite{Georg2}, the agreement with experiment was
poor for two reasons: first, the deconvolved data are artificially rounded;
second, the effect of usual Coulombic interactions have not been taken into
account in the calculation.
\bibitem{asym} The $dI/dV(V)$ are slightly asymetric because the measuring
probe is out of the center of the wire by $0.01L$. Note that the increase of
the interaction rate at intermediate magnetic fields \cite{moriond}, when
the Zeeman splitting of the impurity states is smaller than $eU,$ is not
visible because $f(E)$ is already close to a Fermi distribution at $B=0$ .
This experiment is much more sensitive to the decrease of the exchange rate,
which leads to a sharper structure in $f(E).$
\bibitem{pbjunctions} In practice, we have found that tunnel junctions with
a silver electrode evolve, at room temperature, to an open circuit within a
few hours, hence forbidding implantation of magnetic impurities.
\end{thebibliography}
\end{document}